\documentclass[twocolumn,superscriptaddress,showpacs,aps,amsmath,amssymb,prb]{revtex4-1}
\usepackage{amsfonts}
\usepackage{amsmath}
\usepackage{amssymb}
\usepackage{bm}
\usepackage{graphicx}
\usepackage{color}

\newcommand{\be}{\begin{equation}}
\newcommand{\ee}{\end{equation}}
\newcommand{\bea}{\begin{eqnarray}}
\newcommand{\eea}{\end{eqnarray}}

\begin{document}

\title{Quantum Mechanical Modeling of Nanoscale Light Emitting Diodes}

\author{Rulin Wang}
\affiliation{Beijing Computational Science Research Center, Haidian District, Beijing 100193, China}

\author{Yu Zhang}
\affiliation{Center for Bio-inspired Energy Science, Northwestern University, United States}

\author{Fuzhen Bi}
\affiliation{Beijing Computational Science Research Center, Haidian District, Beijing 100193, China}
\affiliation{University of Science and Technology of China, Hefei, Anhui 230026, China}

\author{GuanHua Chen}
\affiliation{Department of Chemistry, The University of Hong Kong, Pokfulam Road, Hong Kong, China}

\author{ChiYung Yam} \email{yamcy@csrc.ac.cn}
\affiliation{Beijing Computational Science Research Center, Haidian District, Beijing 100193, China}
\affiliation{Department of Chemistry, The University of Hong Kong, Pokfulam Road, Hong Kong, China}

\date{\today}


%
\begin{abstract}
Understanding of the electroluminescence (EL) mechanism in optoelectronic devices
is important for further optimization of their efficiency and effectiveness. Here,
a quantum mechanical approach is formulated for modeling EL processes in nanoscale
light emitting diodes (LED). Based on nonequilibrium Green's function quantum
transport equations, interactions with electromagnetic vacuum environment is
included to describe electrically driven light emission in the devices. Numerical
studies of a silicon nanowire LED device are presented. EL spectra of the nanowire
device under different bias voltages are simulated and, more importantly,
propagation and polarization of emitted photon can be determined using the current
approach.

\end{abstract}

\maketitle
%

Electroluminescence (EL) is an important phenomenon employed in light emitting
diode (LED) technology where light is emitted from a solid state material in
response to an electrical power source. Much work has been devoted to the
development of LED technology that has led to continuous advancements in both
efficiencies and optical power.~\cite{Schubert2005} New efforts are now directed
to exploit semiconductor nanostructures that exhibit extraordinary optical and
electronic properties. A more ambitious use of nanostructure devices is to exploit
quantum effects which fundamentally change the mechanism of electrical-to-optical
power conversion. These devices are made possible with the continuous development
of nanofabrication techniques and are emerging as promising candidates for
optoelectronic and energy devices. Indeed, electrically driven light emission has
been reported from single carbon nanotube and
nanowire~\cite{Mueller2010,Huang2004,Bao2006,Minot2007}, monolayer transition
metal dichalcogenides~\cite{Sundaram2013,Ross2014} and, to the ultimate
miniaturization limit, from a single molecule.~\cite{Marquardt2010,Reecht2014}

Understanding the EL mechanism in nanoscale LED devices is crucial to further
advance the technology for more efficient lighting and enhanced communications.
From the theoretical perspective, accurate description of the
electrical-to-optical conversion processes is a challenging task, since the system
is in nonequilibrium state driven by optical and electric field. In this context,
atomic level modeling is becoming increasingly relevant, not only for accurate
description of the coupled optical-electrical processes, but also to cope with the
myriad of architectures and chemical compositions in modern devices. Prevailing
works evaluate performance of LED devices based on classical models, relying on
parmeters obtained either from experiments~\cite{Kim2007,Malliaras1999} or
first-principles calculations.~\cite{Kordt2015} However, these models fail to
capture quantum phenomena and break down at nanoscale. For microscopic systems,
light emission has been studied using Fermi's golden rule (FGR) to evaluate
transition rates between energy levels.~\cite{Van05236804,Tia11177401,Shi12461}
The first attempt to include quantum effects to simulate directly EL process was
made by Galperin {\it et al.} for model systems.~\cite{Galperin2005,Galperin2012}
Recently, a diagrammatic approach is formulated to study EL in molecular
junctions.~\cite{Harbola2014,Goswami2015} In this letter, we present a quantum
mechanical method for realistic LED device simulations. EL spectra of nanoscale
devices under different bias conditions can be simulated. In addition, the method
offers the possibility of analyzing the polarization of emitted light.

Quantum transport approaches based on nonequilibrium Green's function (NEGF)
method provide an efficient and versatile way to describe the coupled
optical-electrical processes in nanoscale
devices.~\cite{Hen026273,Zhang2014,Meng2015,Yam2015} Based on the Keldysh NEGF
approach, steady state current can be obtained from~\cite{Meir1992}
\begin{eqnarray}
   I_\alpha = \frac{2e}{\hbar} \int \frac{dE}{2\pi}
              {\rm Tr}[\Sigma_\alpha^<(E) G^>(E) - \Sigma_\alpha^>(E) G^<(E)]
\nonumber \\
 \label{current}
\end{eqnarray}
where $G^{<,>}$ are lesser and greater Green's functions, providing information on
the energy states and population statistics for electrons and holes, respectively.
$\Sigma^{<,>}_{\alpha}$ are the self-energies and $\alpha$ corresponds to a
particular scattering process. Considering a two-terminal LED device, the
scattering processes arise from the contacts and also electron-photon interaction.
The first and second terms in square bracket of Eq.~(\ref{current}) are
interpreted respectively as the incoming and outgoing rate of electrons in device
due to the scattering processes. Thus, $I_\alpha$ gives the steady state current
resulting from different scattering processes.

The self-energy associated to the contacts can be obtained following standard
procedure,~\cite{Xue2002} whereas the explicit evaluation of electron-photon
self-energy, $\Sigma^{<,>}_{\rm ep}$ requires many body diagrammatic technique and
its self-consistent Born Approximation (SCBA) expression is given by~\cite{Frederiksen2007,Zhang2013}
\begin{eqnarray}
  \Sigma^{<,>}_{\rm ep} (E) = & \sum_{q} M_{q}
                                \left[
                                (N_{q} + 1) G^{<,>}(E \pm \hbar\omega_{q})
                                \right.
\nonumber \\
                              & \left.
                              + N_{q} G^{<,>}(E \mp \hbar\omega_{q})
                                \right] M_{q}
  \label{selfenergy}
\end{eqnarray}
where $N_{q}$ is photon occupation number and $\omega_{q}$ is photon frequency.
$q$ refers to photon mode characterized by its wave vector $\vec{k}_q$ and
polarization directions $\vec{\lambda}_q$. The three vectors are mutually
perpendicular with each other and are defined as
\begin{eqnarray}
  \left\{ \begin{aligned}
    &\vec{k} = (\sin\theta \cos \phi, \sin\theta \sin\phi, \cos\theta)  \\
    &\vec{\lambda}_{q,\|}   = (\sin \phi, - \cos \phi, 0) \\
    &\vec{\lambda}_{q,\bot} = (\cos \theta \cos \phi, \cos \theta \sin \phi, -\sin \theta)
    \end{aligned} \right.
\end{eqnarray}
For EL processes, the associated self-energy accounts for interactions with
electromagnetic field modes in their vacuum state ($N_{q}$ = 0). The system then
undergoes spontaneous emission by relaxation to a lower energy state.
$\Sigma^{<,>}_{\rm ep}$ for spontaneous emission is thus given by
\begin{eqnarray}
  \Sigma^{<,>}_{\rm ep}(E) = & \sum_{q} M_{q}
                               G^{<,>}(E \pm \hbar\omega_{q}) M_{q}
\end{eqnarray}
Here, $M_q$ is electron-photon coupling matrix and its elements are given by~\cite{Hen026273,Zhang2014}
\begin{eqnarray}
   M_{q,\mu\nu} = \frac{e}{m}(\frac{\hbar}{2 \epsilon_0 \omega_{q} V})^{1/2}
                  \vec{\lambda}_{q} \cdot \langle \mu | \vec{p} | \nu \rangle
\end{eqnarray}
Here, $\hbar$ is reduced Planck constant; $\epsilon_0$ is vacuum permittivity; $V$ is volume.
The infinite sum in Eq.~(\ref{selfenergy}) is tranformed to integration
\begin{eqnarray}
  \Sigma^{<,>}_{\rm ep}(E) = \int_0^\infty d(\hbar \omega)
                             \int_0^\pi d\theta \sin \theta
                             \int_0^{2\pi} d\phi
\nonumber \\
                           \times \left[
                             \Sigma^{<,>}_{\|}(E,\omega,\theta,\phi)
                           + \Sigma^{<,>}_{\bot}(E,\omega,\theta,\phi)
                           \right]
  \label{selfenergy2}
\end{eqnarray}
where $\Sigma^{<,>}_{\|}(E,\omega,\theta,\phi)$ and
$\Sigma^{<,>}_{\bot}(E,\omega,\theta,\phi)$ are defined as angle-dispersed
self-energies for the two perpendicular polarization directions,
\begin{eqnarray}
  \Sigma^{<,>}_{\|}(E,\omega,\theta,\phi) = & R^{<,>}_{xx} \sin^2 \phi + R^{<,>}_{yy} \cos^2 \phi
\nonumber \\
                                            & - (R^{<,>}_{xy} + R^{<,>}_{yx}) \sin \phi \cos \phi
  \label{selfenergy3}
\end{eqnarray}
\begin{eqnarray}
  \Sigma^{<,>}_{\bot}(E,\omega,\theta,\phi) &=& \big[
                                                  R^{<,>}_{xx} \cos^2 \phi + (R^{<,>}_{xy} + R^{<,>}_{yx}) \cos \phi \sin \phi
\nonumber \\
                                            & & + R^{<,>}_{yy} \sin^2 \phi \big]\cos^2 \theta + R_{zz}^{<,>} \sin^2 \theta
\nonumber \\
                                            & & - (R^{<,>}_{xz} + R^{<,>}_{zx}) \cos \phi \cos \theta \sin \theta
\nonumber \\
                                            & & - (R^{<,>}_{yz} + R^{<,>}_{zy}) \sin \phi \cos \theta \sin \theta
  \label{selfenergy4}
\end{eqnarray}
and
\begin{eqnarray}
   R_{ij}^{<,>} &=& P_i G^{<,>}(E \pm \hbar \omega) P_j
\nonumber \\
   P_{i,\mu\nu} &=& (\frac{\omega e^2 }{16 \pi^3 c^3 m^2 \epsilon_0})^{1/2} \langle \mu |p_i| \nu \rangle
\end{eqnarray}
and $i,j \in (x,y,z)$. The Green's function in Eq.~(\ref{current}) can then be
obtained from the Keldysh equation
\begin{eqnarray}
  G^{<,>}(E) = \sum_\alpha G^r(E) \Sigma^{<,>}_{\alpha}(E) G^a(E).
\end{eqnarray}
where $G^r$ and $G^a$ are retarded and advanced Green's functions.

Substituting Eq.~(\ref{selfenergy2}) into Eq.~(\ref{current}), $I_{\rm ep}$ in
Eq.~(\ref{current}) should be zero since number of electrons should be conserved
during emission of photons. Thus, the first term in Eq.~(\ref{current})
corresponds to transition of electron from energy level $E + \hbar\omega$ to $E$
while emitting a photon with energy $\hbar\omega$. And the emission flux
$F^{\rm em}$ for photon frequency $\omega$ can be obtained by
\begin{align}
  F^{\rm em}(\omega)= \frac{2}{\hbar} \int \frac{dE}{2\pi} {\rm Tr}[\Sigma_{\rm ep}^<(E) G^>(E)]
  \label{emission}
\end{align}
More importantly, the wave vector and polarization of emitted photons can be
determined by substituting the angle-dispersed self-energies
Eqs.~(\ref{selfenergy3}) and (\ref{selfenergy4}) into Eq.~(\ref{emission}).
\begin{align}
  F^{\rm em}_{\|}(\omega,\theta,\phi) = \frac{2}{\hbar} \int \frac{dE}{2\pi}
                               {\rm Tr}[\Sigma_{\|}^<(E,\omega,\theta,\phi) G^>(E)]
\nonumber \\
  F^{\rm em}_{\bot}(\omega,\theta,\phi) = \frac{2}{\hbar} \int \frac{dE}{2\pi}
                                  {\rm Tr}[\Sigma_{\bot}^<(E,\omega,\theta,\phi) G^>(E)]
  \label{emissionangle}
\end{align}
FGR has been commonly used to evaluate rate of spontaneous emissions. For simple
two-level systems, Eq.~(\ref{emission}) recovers FGR rate expresson for electron
transition between the levels. It is
important to emphasize that the current approach offers the possibility to
determine the polarization of emitted photons and describe the nonequilibrium
statistics of the device due to the bias voltage and interactions with photons.

\begin{figure}[ht]
  \includegraphics[width=80mm]{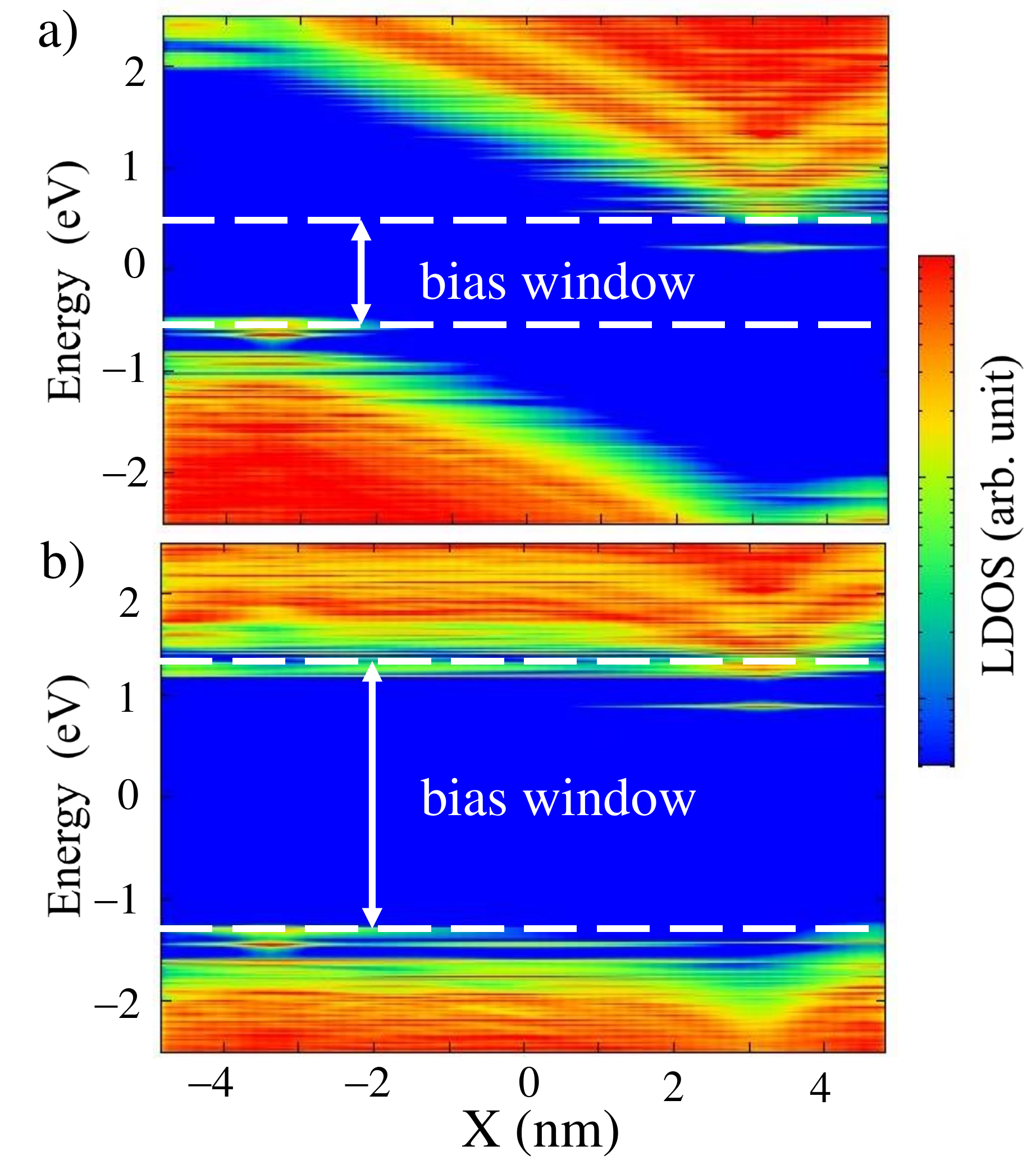}
  \caption{ LDOS of device along the nanowire axis for forward bias voltage of
            (a) 1.0 V and (b) 2.6 V. The left side of nanowire is $p$-doped and
            right side is $n$-doped. A built-in potential is formed at the
            junction due to the space charge.}
  \label{LDOS}
\end{figure}

We apply the method to model a nanoscale LED device based on a Si nanowire with
cross section diameter of 1.5 nm. The nanowire is 9.5 nm in length oriented in
[110] direction. Atomistic model is employed in current study which contains 1000
atoms. To form a $p$-$n$ junction, Ga and As atoms are explicitly doped in the
system to give a doping concentration of about $2.0 \times 10^{20} {\rm cm}^{-3}$.
The surface of nanowire is passivated with hydrogen atoms to eliminate dangling
bonds. The device is connected to two semi-infinite doped Si leads where external
bias voltage is applied. The electronic structure of the model is described at the
density functional tight-binding (DFTB) level.~\cite{Por9512947,Els987260} At
equilibrium, an internal built-in voltage $V_{\rm bi}$ of 2.44 V is formed across
the two different doped regions. The simulations are performed at 300 K.

\begin{figure}[ht]
  \includegraphics[width=85mm]{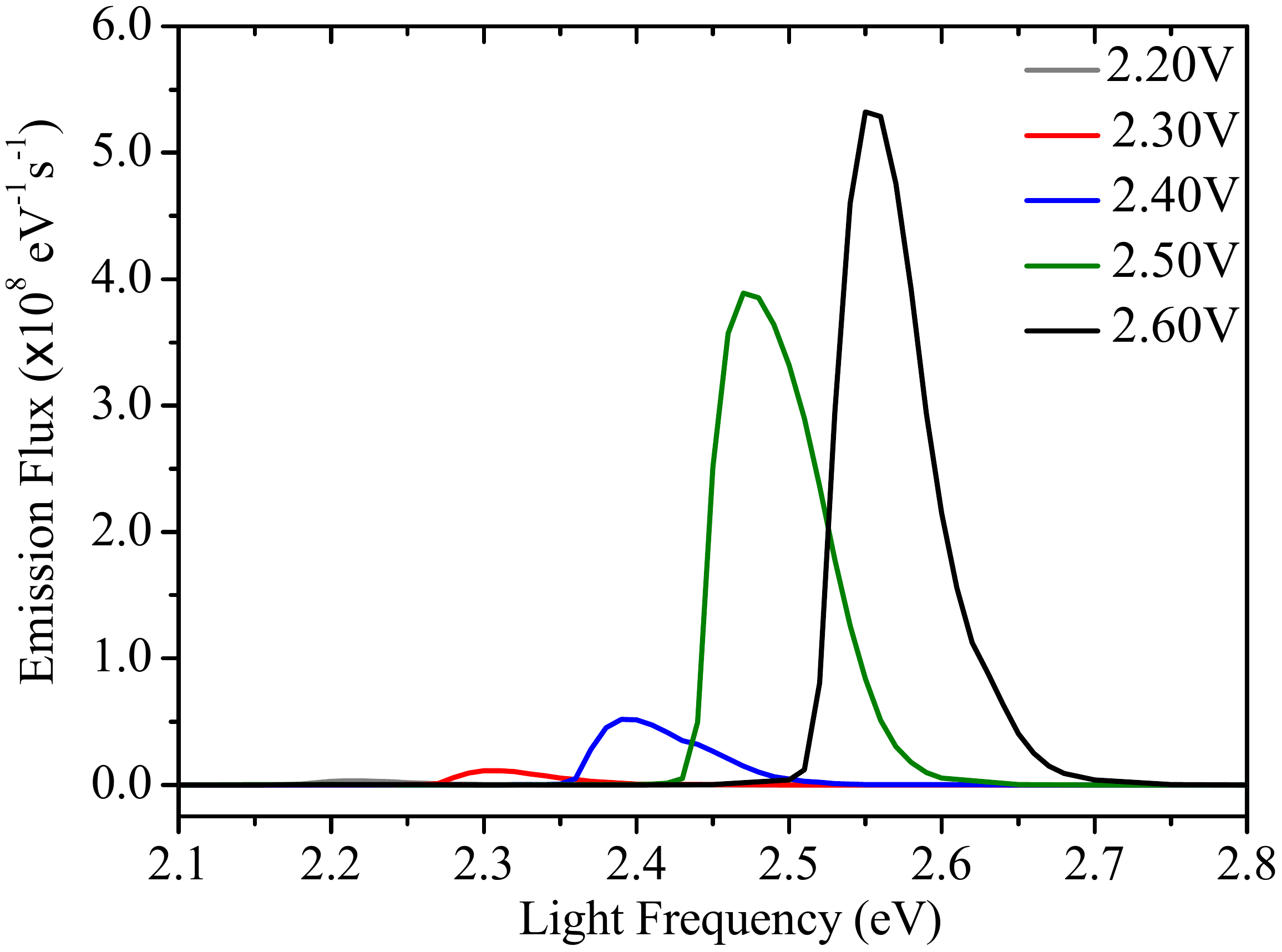}
  \caption{Electroluminescence spectrum of the Si nanowire LED device for various
           forward bias voltages. Grey line: 2.2 V; Red line: 2.3 V; Blue line:
           2.4 V; Green line: 2.5 V; Black line: 2.6 V.}
  \label{elspectra}
\end{figure}

We solve Eq.~(\ref{emission}) to obtain EL spectra of the nanowire device under
different external bias voltage. In this work, the lowest order expansion to the
self-energy $\Sigma^{<,>}_{\rm ep}$ is employed. Physically, this corresponds to
the situation where density of states (DOS) of the device is unaffected by
electron-photon interaction. This can be justified by the fact that interaction
with electromagnetic vacuum environment is weak. Therefore, electronic structure
remains intact and nonlinear effects are neglected. Fig.~\ref{LDOS} plots the
local density of states (LDOS) of the device along the wire direction for forward
bias voltages of (a) 1.0 V and (b) 2.6 V. Clearly, a built-in voltage is formed
across the junction, as shown in Fig.~\ref{LDOS}(a). Due to this potential
barrier, electrons are localized at the $n$-doped region while holes are localized
at the $p$-doped region. The electron-hole recombination is inhibited and the
emission process is suppressed in this case. When the forward bias is increased,
the potential difference across the junction is reduced. As shown in
Fig.~\ref{LDOS}(b), conducting channels are formed at conduction band and valence
band edges for electrons and holes, respectively. The carriers can then move along
the channels driven by the external bias voltage. Due to their spatial proximity,
the electron-hole pairs undergo a recombination and energy is emitted in form of
photons.

\begin{figure}[ht]
  \includegraphics[width=75mm]{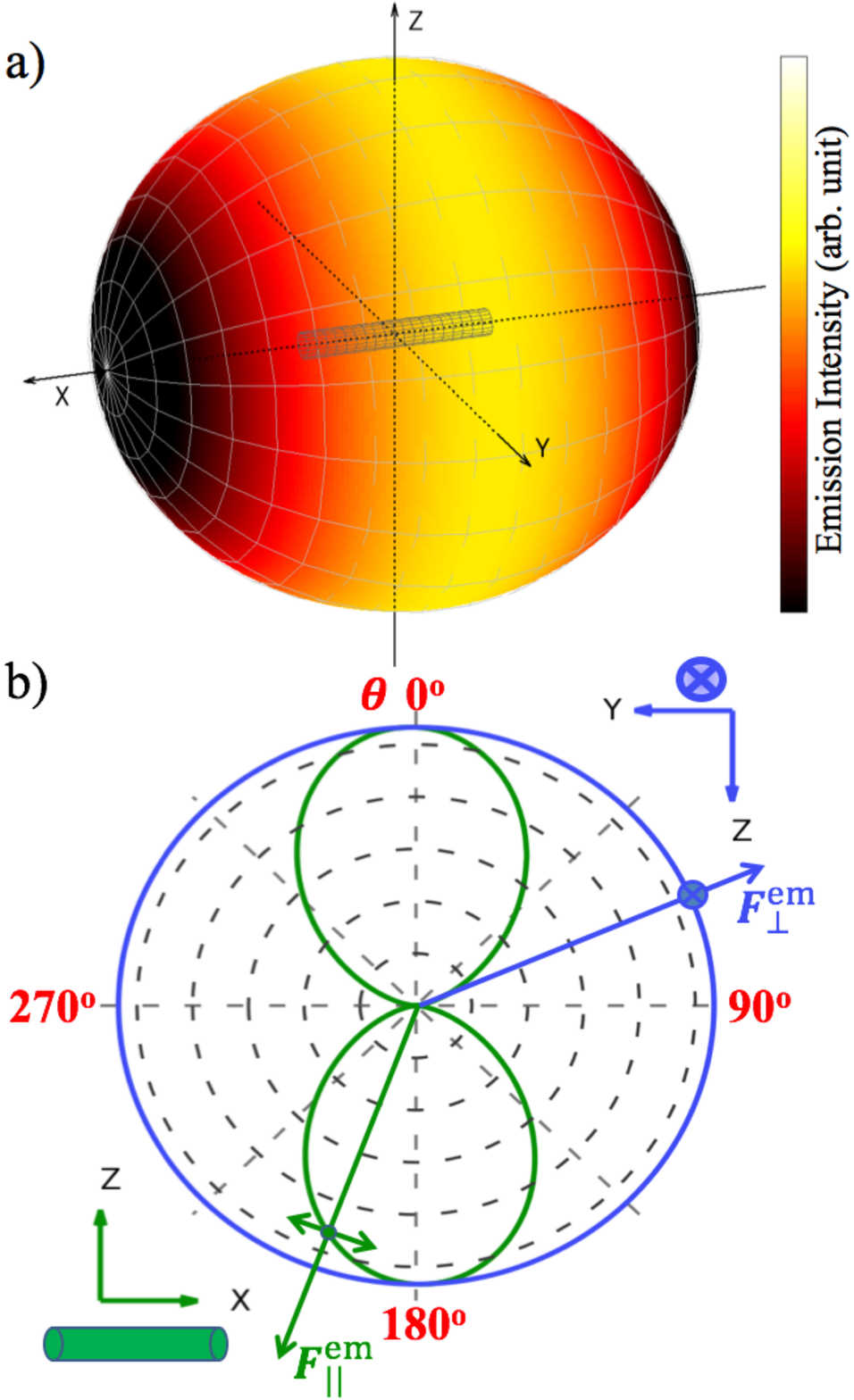}
  \caption{ (a) EL intensity distribution of the Si LED device under 2.4 V forward
            bias voltage. Light frequency is set as 2.4 eV. The color represents
            intensity of emitted light and angular coordinates correspond to the
            propagation direction. (b) Polar plot of EL intensity as a function
            of $\theta$. Green line: $F^{\rm em}_\|$ in the $x-z$ plane; Blue
            line: $F^{\rm em}_\bot$ in the $y-z$ plane.}
  \label{polar}
\end{figure}

EL spectra of the nanowire LED device is plotted in Fig.~\ref{elspectra} for
different bias voltage. A single broad emission peak is observed corresponding to
transitions from conduction band to valence band. This is in contrast to that of
molecular junctions~\cite{Goswami2015} where multiple peaks are observed due to
molecular resonances. The shape of emission peak is asymmetric with tail at
higher energy side due to the Fermi-Dirac distribution of charge carriers. We note
that the intensity of photon emission in general increases with applied bias
voltage. For bias voltage below 2.0 V, no light emission is observed. This is
consistent with the results shown in the LDOS, where electron-hole recombination
is suppressed when applied bias voltage is lower than the internal built-in
voltage of the device. As the forward bias approaches flat band position,
electrons and holes are injected simultaneously from electrodes and recombine at
the junction where they meet. The emission intensity therefore increases
substantially when the applied bias exceeds the built-in potential of the system,
as shown in Fig.~\ref{elspectra}. For bias volatge of 2.6 V, a strong EL peak at
light frequency of 2.55 eV is observed. In general, charge carriers relax
nonradiatively as they pass through the device and results in near band edge
emission. The system studied in this work is small compared to the coherence
length.~\cite{Lu2005} Electron-phonon interactions are thus neglected in the
simulations and inelastic scatterings are assumed to be caused only by photons.
Phonon scattering can be included similarly as Eq.~(\ref{selfenergy}) within NEGF
formalism~\cite{Galperin2007,Pecchia2007,Dubi2011} and its effect on EL of
nanoscale device needs further investigations.

The optical emission from the nanowire LED device is further characterized by its
propagation and polarization. Eq.~(\ref{emissionangle}) allows analysis of its
spatial distribution along the two polarization vectors. Fig.~\ref{polar}(a) shows
the EL intensity distribution of the Si LED device under bias voltage of 2.4 V.
Emitted light frequency is chosen as 2.4 eV. The Si nanowire is oriented along
$x$-axis. The key features we note in Fig.~\ref{polar}(a) are that light is
emitted mainly from surface of nanowire and essentially no edge emission is
observed. We further analyse the polarization of emitted light in
Fig.~\ref{polar}(b). The green line gives the polar plot of the emission flux
$F^{\rm em}_\|$ in the $x-z$ plane while blue line plots $F^{\rm em}_\bot$ in the
$y-z$ plane. Here, $\theta$ is defined as the angle measured from $z$-axis.
$F^{\rm em}_\|$ represents the in-plane polarization which makes an angle $\theta$
with respect to the nanowire axis. As shown in Fig.~\ref{polar}(b),
$F^{\rm em}_\|$ (green line) is proportional to $\cos^2\theta$, giving maximum EL
intensity when it is aligned parallel to the nanowire axis. $F^{\rm em}_\bot$
(blue line) represents the out-of-plane polarization and is always aligned parallel
to the nanowire axis. Thus, $F^{\rm em}_\bot$ in $y-z$ plane remains
constant with respect to $\theta$. Our results clearly show that the Si nanowire
LED behaves as a linearly polarized radiation source. This is consistent with
experimental observation of light emission from a carbon nanotube
device.~\cite{Misewich2003}

In conclusion, we formulate a quantum mechanical approach for modeling nanoscale
LED devices based on NEGF quantum transport formalism. The nonequilibrium
statistics in the device due to applied voltage and interactions with light are
taken into account and EL processes in LED devices can be accurately described.
The current approach provides the tools for determining not only the intensity but
also propagation and polarization of optical emission in nanoscale devices. We
demonstrate the method by simulations of EL properties of a Si nanowire LED
device. Given the complexity of modern nanoscale devices, atomistic details and
quantum effects are playing increasingly important roles in determining the device
properties. Important also is to understand EL of single molecules in scanning
tunneling microscopy experiments.~\cite{Berndt1993,Wu2006} The quantum mechanical
method presented in this work provides an efficient research tool for theoretical
studies of coupled optical-electrical processes in these nanoscale systems.

\acknowledgments
The authors would like to thank Wen Yang for helpful discussions. The financial
support from the National Natural Science Foundation of China (21322306(C.Y.Y.),
21273186(G.H.C., C.Y.Y.)), National Basic Research Program of China
(No. 2014CB921402 (C.Y.Y.)), and University Grant Council
(AoE/P-04/08(G.H.C., C.Y.Y.)) is gratefully acknowledged.



%
%

\bibliography{main}

%
%

\end{document}